\documentclass[12pt,preprint]{aastex}  

\usepackage{epsfig}
\usepackage{natbib}
\usepackage{slashbox}

\usepackage{CJK}

\shorttitle{Flux Rope Eruptions for Magnetar Giant Flares}

\shortauthors{}

\begin{document}

\title{Magnetar Giant Flares --- Flux Rope Eruptions \\
in Multipolar Magnetospheric Magnetic Fields}

\author{Cong ~Yu\altaffilmark{1,2}}
\altaffiltext{1}{National Astronomical Observatories/Yunnan
Astronomical Observatory, Chinese Academy of Sciences, Kunming,
650011, China; {\tt cyu@ynao.ac.cn}}

\altaffiltext{2}{Key Laboratory for the Structure and Evolution of
Celestial Objects, Chinese Academy of Sciences, Kunming, 650011,
China}

\date{$\qquad\qquad\qquad\qquad\qquad \qquad\qquad$ July 23 2012}


\begin{abstract}
We address a primary question regarding the physical mechanism
that triggers the energy release and initiates the onset of
eruptions in the magnetar magnetosphere. A self-consistent
stationary, axisymmetric model of the magnetar magnetosphere is
constructed based on a force-free magnetic field configuration
which contains a helically twisted force-free flux rope. The
magnetic field configurations in the magnetosphere are obtained as
solutions of an inhomogeneous Grad-Shafranov (GS) equation. Given
the complex multipolar magnetic fields at the magnetar surface, we
also develop a convenient numerical scheme to solve the GS
equation. Depending on the surface magnetic field polarity, there
exist two kinds of magnetic field configurations, inverse and
normal. For these two kinds of configurations, variations of the
flux rope equilibrium height in response to gradual surface
physical processes, such as flux injections and crust motions, are
carefully examined. We find that equilibrium curves contain two
branches, one represents a stable equilibrium branch, the other an
unstable equilibrium branch. As a result, the evolution of the
system shows a catastrophic behavior: when the magnetar surface
magnetic field evolves slowly, the height of flux rope would
gradually reach a critical value beyond which stable equilibriums
can no longer be maintained. Subsequently the flux rope would lose
equilibrium and the gradual quasi-static evolution of the magnetar
magnetosphere will be replaced by a fast dynamical evolution. In
addition to flux injections, the relative motion of active regions
would give rise to the catastrophic behavior and lead to magnetic
eruptions as well. We propose that a gradual process could lead to
a sudden release of magnetosphere energy on a very short dynamical
timescale, without being initiated by a sudden fracture in the
crust of the magnetar. Some implications of our model are also
discussed.


\end{abstract}



\section{INTRODUCTION}
Two intimately connected classes of young neutron stars $-$ Soft
Gamma-ray Repeaters (SGRs) and Anomalous X-ray Pulsars (AXPs),
which are commonly referred to as magnetars, both show high energy
emissions (Mazets et al. 1979; Mereghetti \& Stella 1995;
Kouveliotou et al. 1998; Gavriil et al. 2002). It is widely
believed that the X-ray luminosity in these sources is powered by
the dissipation of non-potential (current-carrying) magnetic
fields in the ultra-strongly magnetized magnetosphere with the
magnetic field $\mathbf{B}$ $\sim 10^{14}-10^{15}$G (Duncan \&
Thompson 1992; Thompson \& Duncan 1996; Thompson et al. 2002).
Occasionally, a much brighter outburst has been observed, i.e., a
giant flare releases a total energy of $\sim 10^{46}$ergs and has
a peak luminosity of $\sim 10^{44 - 46}$ergs $\mathrm{s}^{-1}$
(for recent reviews see Woods \& Thompson 2006; Mereghetti 2008).
Although the energy for magnetar outbursts is widely believed to
be supplied by the star's magnetic field, the physical process by
which the energy is stored and released remains one of the great
puzzles in high-energy astrophysics. Two possibilities exist for
the location where the magnetic energy is stored prior to an
eruption: in the magnetar crust or in the magnetosphere. For the
former possibility, a giant flare may be caused by a sudden
untwisting of the internal (to the neutron star) magnetic field
(Thompson \& Duncan 2001). The subsequent quick and brittle
fracture of the crust leads to energetic
outbursts\footnote{However, recent calculations by Levin \&
Lyutikov (2012) imply that plastic deformations of the crust are
more likely to occur and the crust model of giant flares may not
explain the fast dynamical energy release.}. During the outbursts,
there would be an enhanced twist of the magnetospheric magnetic
field lines. In this crust scenario, the energy stored in the
external twist is limited by the tensile strength of the crust.
Alternatively, due to the difficulties to explain the short
timescale of the giant flare rise time, $\sim$ 0.25 ms (Palmer et
al. 2005), the second possibility --- the magnetospheric storage
model, was proposed by Lyutikov (2006). In this particular
scenario, the energy stored in the external twist need not be
limited by the tensile strength of the crust, but instead by the
total external magnetic field energy (Yu 2011b). In the
magnetospheric storage model, the magnetic energy storage
processes take place quasi-statically on a longer timescale than
the dynamical flare timescale prior to the eruption.

In the magnetospheric model for giant flares, the energy released
during an eruption is built up gradually in the magnetosphere
before the eruption. Some interesting properties about the storage
of magnetic energy of the magnetospheric models have been
discussed in Yu (2011b). But there still remains a primary
question regarding magnetospheric models, i.e., 
what is the mechanism that triggers the energy release and
initiates the eruption? More specifically, how a very gradual
process by the flux injections (Klu\'{z}niak \& Ruderman 1998;
Thompson et al. 2002) or crust motions (Ruderman 1991) could lead
to a sudden release of magnetosphere energy on a very short
dynamical timescale, without being initiated by a sudden fracture
in the rigid component of the neutron star.
This catastrophic behavior is essentially reminiscent of solar
flares and coronal mass ejections (CMEs). It is conceivable that
the magnetosphere adjusts quasi-statically in response to the
slowly-changing boundary conditions at the magnetar surface. After
reaching a critical point, the magnetosphere could no longer
maintain a stable equilibrium and a sudden reconfiguration of the
magnetic field occurs due to loss of equilibrium (Forbes \&
Isenberg 1991; Isenberg et al. 1993; Forbes \& Priest 1995). The
subsequent physical processes would proceed on a dynamical time
scale. This catastrophic process naturally explains the puzzle how
a very slow process could lead to the sudden release of external
magnetic energy on a much shorter timescale (Thompson et al.
2002).
The magnetar giant flares may involve a sudden loss of equilibrium
in the magnetosphere, in close analogy to solar flares and CMEs
(Lyutikov 2006).
A number of CMEs show structures consistent with the ejection of a
magnetic flux rope\footnote{The flux rope is a helically twisted
magnetic arcade anchored on the solar surface and often used to
model prominences in the solar corona.}, as has been reported by
Chen et al.(1997) and Dere et al. (1999). Hence magnetic flux
ropes have been presumed to be typical structures in the solar
corona, and their eruptions might be closely related to solar
flares and CMEs (Forbes \& Isenberg 1991; Isenberg et al. 1993).
Similarly, in the magnetar magnetosphere, magnetic flux ropes
could be generated due to the pre-flare activity (G\"{o}tz et al.
2007; Gill \& Heyl 2010). As the magnetic flux injects from deep
inside the magnetar, the dissipation of the magnetic field may
give rise to the precursor activity. The magnetic dissipation of
the precursor could also lead to topology changes of the magnetic
fields and the formation of a magnetic flux rope\footnote{In this
work, we do not address the question of how a flux rope might be
formed. Possible mechanism had been discussed by van Ballegooijen
\& Martens (1989).}. Such a flux rope is also an indispensable
ingredient for the radio afterglow observed in SGR1806 (Gaensler
et al. 2005; Lyutikov 2006). It is worthwhile to note that the
magnetic field interior to the flux rope, which is suspended in
the magnetosphere, is helically twisted. It corresponds to a
locally twisted feature in the magnetosphere (Thompson et al 2002;
Pavan et al. 2009). Such locally twisted flux ropes seem to be
more consistent with recent observations, which suggest the
presence of localized twist, rather than global twist (e.g. Woods
et al. 2007; Perna \& Gotthelf 2008).

Observations show a striking feature that the emergence of a
strong four-peaked pattern in the light curve of the 1998 August
27 event from SGR 1900+14, which was shown in data from the
Ulysses and BeppoSAX gamma-ray detectors (Feroci et al. 2001).
These remarkable data may imply that the geometry of the magnetic
field was quite complicated in regions close to the star. It is
reasonable to infer that, near the magnetar surface, the magnetic
field geometry of an SGR/AXP source involves higher multi-poles.
The multipolar magnetic field configurations could be readily
understood within the magnetar model. Physically speaking, the
electric currents, formed during the birth of magnetar, slowly
push out from within the magnetar and generate active regions on
the magnetar surface. These active regions manifest themselves as
the multipolar regions on the magnetar surface. Due to the
presence of the active regions, the magnetic field may deviates
from a simple dipole configuration near the magnetar surface
(Pavan et al. 2009). Our calculations show that multipolar
magnetic active regions, especially their relative motions, would
have important implications for the catastrophic eruptions of
magnetar giant flares (see Section 4).

Motivated by the similarity between giant flares and solar CMEs,
Lyutikov (2006) speculated that magnetar giant flares may also be
trigged by the loss of equilibrium of a magnetic field containing
a twisted flux rope. But no solid calculations about the
equilibrium loss of a flux rope in magnetar magnetosphere have
been performed yet. In this paper we focus on the possibility of
magnetospheric origin for giant flares and propose that the
gradual variations at the magnetar surface could lead to fast
dynamical processes in the magnetosphere. We will construct a
force-free magnetosphere model with a flux rope suspended in the
magnetosphere and study the catastrophic behavior of the flux rope
in a background multi-polar magnetic field configuration, taking
into account the possible effects of flux injections (Klu\'{z}niak
\& Ruderman 1998; Thompson et al. 2002) and crust horizontal
motions (Ruderman 1991). We are especially interested in the
critical height of the flux rope that can be achieved in our
model.
In the mean time, we also develop a convenient numerical scheme to
solve the inhomogeneous Grad-Shafranov (GS) equation. Since
observed magnetars have a very slow rotation rate, we ignore
rotation effects throughout this work.

This paper is structured as follows: in Section 2 we describe the
basic equations for the force-free magnetosphere model as well as
the multipolar boundary conditions. Two possible magnetic
configurations are also discussed in this section. In Section 3 we
will discuss the internal and external equilibrium constraints in
our model. Numerical results about catastrophic behaviors of the
magnetosphere in response to flux injections and crust motions are
discussed in Section 4. Conclusions and discussions are given in
Section 5. Technical details about the force-free magnetosphere
magnetic field are given in Appendix A and B.

\section{Axisymmetric Force-Free Magnetosphere with Multipolar Boundary Conditions}
The magnetic fields in the magnetar magnetosphere are so strong
that the inertia and pressure of the plasma could be ignored
(Thompson et al. 2002; Yu 2011a). As a result, the magnetosphere
is assumed to be in a force-free equilibrium state, in which
$\mathbf{J}\times\mathbf{B} = 0$.
The axisymmetric force-free magnetic field configurations is
determined by an inhomogeneous Grad-Shafranov (GS) equation.
Throughout this paper, we work in the spherical polar coordinates
($r,\theta, \phi$).

\subsection{Force-Free Magnetic Field Containing a Flux Rope }
In our magnetosphere model, one of the distinguishing features is
that there exists a helically twisted flux rope in the
magnetosphere. The precursor of a giant flare could be relevant to
the formation of such helically twisted flux ropes (G\"{o}tz et
al. 2007; Gill \& Heyl 2010). Due to the presence of the flux
rope, the magnetic fields consist of two parts, one is the fields
that are inside the flux rope and the other is the fields outside
the flux rope.

In our model the magnetic twist of the flux rope is locally
confined to the flux rope interior. This is quite different from
the globally twisted magnetic field configurations in Thompson et
al. (2002) and Beloborodov (2009). They considered a non-potential
force-free field where the electric currents permeate through the
entire magnetosphere, while our model only contains a electric
current in a spatially confined region, i.e., interior to the flux
rope. Note that the toroidal flux rope has a minor radius, $r_0$,
which is small compared to the height of flux rope, $h$, which is
actually the major radius of the flux rope. Under such
circumstances, the magnetic field produced by the current inside
the flux rope can be viewed as that produced by a wire carrying
the net current $I$ at the center of the flux rope (Forbes \&
Priest 1991) and a simple Lundquist (1950) force-free solution
could be applied to represent the distribution of current density
and magnetic field inside the flux rope. A brief yet
self-contained description of the Lundquist solution is given in
Appendix A.

Outside the flux rope, the magnetic field is essentially
potential, i.e., the field outside the flux rope is non-twisting.
In the regions exterior to the flux rope, the steady state
axisymmetric magnetic field in the magnetosphere has only poloidal
components and can be written as
\begin{equation}
{\bf B} = 
\nabla \Psi \times \nabla\phi \ ,
\end{equation}
where $\Psi(r,\theta)$ is the magnetic stream function and $\phi$
is the third component of the spherical polar coordinates. Written
explicitly, the magnetic field is
\begin{equation}\label{brbt}
{\bf B} = \frac{1}{r\sin\theta} \left( \frac{1}{r}\frac{\partial
\Psi}{\partial \theta} , \  - \frac{\partial \Psi}{\partial
r} \right) \ . 
\end{equation}
The force-free condition can be cast into the standard
Grad-Shafranov (GS) equation (Thompson et al. 2002)
\begin{equation}\label{inhomoGS}
\frac{\partial^2 \Psi}{\partial r^2} %
+ \frac{\sin\theta}{r^2}\frac{\partial}{\partial \theta}%
\left( \frac{1}{\sin\theta}\frac{\partial \Psi}{\partial\theta} \right) %
 = - (r\sin\theta) \frac{4\pi}{c} J_{\phi} \ ,
\end{equation}
where $c$ is the speed of light. The current density $J_{\phi}$ on
the right hand side of the above equation is caused by a toroidal
force-free magnetic flux rope mentioned above, which is suspended
in the magnetosphere at the height, $h$, by force balances. Note
that the stream function $\Psi$ is determined simultaneously by
the electric current inside the flux rope and boundary conditions
at the magnetar surface. The boundary conditions will be discussed
separately in next section.
The electric current inside the flux rope could be treated as a
source term on the right hand side of the above inhomogeneous GS
equation. We treat the current density of the flux rope as a
circular ring current of the form (Priest \& Forbes 2000)
\begin{equation}\label{currentdensity}
J_{\phi} = \frac{I}{h} \ \delta(\cos\theta)\delta(r-h) \ ,
\end{equation}
where $I$ is the electric current carried by the flux rope.
Similar treatments have been adopted in the coronal mass ejection
(CME) studies (Forbes et al. 1991; Lin et al. 1998). It is clear
from this equation that the flux rope is located at the equatorial
plane ($\theta = \pi/2$) and the flux rope is the only current
source in the region $r>r_s$, where $r_s$ is the magnetar radius
(also see Fig. 2).




\subsection{Multipolar Boundary Conditions at Magnetar Surface }
In order to solve the boundary-value problem associated with the
inhomogeneous GS equation (\ref{inhomoGS}), we still need to know
the boundary condition at the magnetar surface $r = r_s$ (where
$r_s$ is the magnetar radius)\footnote{The boundary condition at
$r \rightarrow\infty$ is simply $|\nabla \Psi| \rightarrow 0 $,
which is satisfied trivially in this work, see Appendix B.}.
We choose $\Psi$ at the magnetar surface $r = r_s$ to be
\begin{equation}\label{BCs}
\Psi_s (r_{s}, \mu) = \Psi_0 \sigma \Theta (\mu) \ ,
\end{equation}
where the subscript $s$ denotes the value on the neutron star
surface, $\Psi_0$ is a constant with magnetic flux dimension,
$\sigma$ is a dimensionless quantity that determines the magnitude
of the flux at the surface, and  $\mu = \cos\theta$. The magnetic
field configuration of neutron stars is basically a dipole field.
But near the neutron star surface, where the loss of equilibrium
occurs, the magnetic field presents much more complex behaviors
(Feroci et al. 2001). To simulate multipolar regions on the
neutron star surface, we add two Gaussian functions to the usual
dipole field and consequently the function $\Theta(\mu)$ in the
above equation takes the following form,
\begin{equation}\label{thetafunction}
\Theta(\mu) \equiv  (1-\mu^2) + %
\exp\left[-\frac{(\mu-\mu_0)^2}{2 w^2}\right] + %
\exp\left[-\frac{(\mu+\mu_0)^2}{2 w^2}  \right] \ ,
\end{equation}
where $\mu_0$ and $w$ are parameters that determine the magnetic
flux distributions at the neutron star surface. We take $w=0.001$
throughout this paper. It is worthwhile to note 
that, according to the parameter $\Psi_0$ introduced in Equation
(\ref{BCs}), a dimensional current can be defined as
\begin{equation}
I_0 = \frac{\Psi_0 c}{r_s} \ ,
\end{equation}
where $r_s$ is the radius of the magnetar. Throughout this paper,
we scale all lengths by the neutron star radius $r_s$, magnetic
flux by $\Psi_0$ and current by $I_0 = \Psi_0 c /r_s$. We also
define a dimensionless current $J = I/I_0$ for later use, where
$I$ is the electric current carried by the flux rope (see Section
2.1).

\begin{figure}
\begin{center}
\epsfig{file=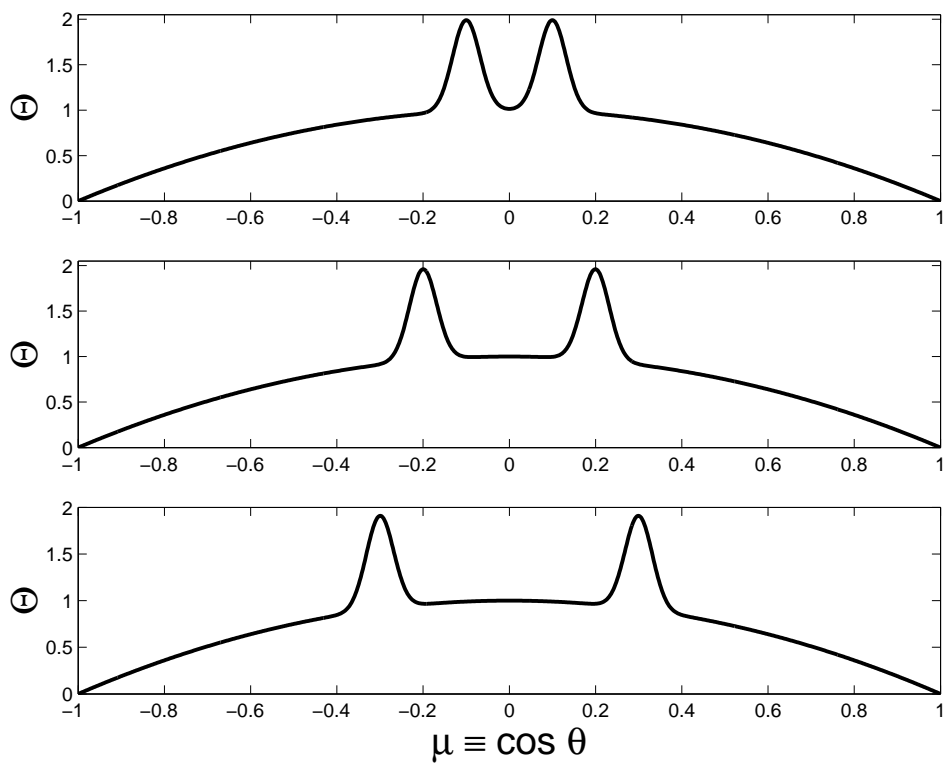,height=4in,width=5in,angle=0}
\end{center}
\caption{
         Distribution of magnetic flux at the magnetar
         surface. Three panels correspond to $\mu_0 = 0.1$,
         $\mu_0 = 0.2$, and $\mu_0 = 0.3$, respectively. The
         increase of $\mu_0$ clearly shows that the magnetic
         active regions are moving apart, which will have
         significant implications for the catastrophic behavior
         of flux rope.
         }
\label{fig1}
\end{figure}

Fig. \ref{fig1} shows the profile of the flux function
$\Theta(\mu)$ in Equation (\ref{thetafunction}). Note that the
derivative with respect to $\mu$ gives the radial component of
magnetic field at the magnetar surface. This boundary flux
distribution is symmetric with respect to the equator $\theta =
\pi/2$ or $\mu = 0$.   In real circumstances, the active regions
on magnetars may form much more complicated patterns without any
symmetry. For simplicity, we focus in this paper only on systems
with such symmetry. The distance between the two active regions is
specified by the parameter $\mu_0$.
The distribution of $\Theta(\mu)$ for three different values of
$\mu_0 = 0.1, 0.2,$ and $0.3$ are shown in three panels of Fig. 1.
It is clear from this figure that, with the increase of $\mu_0$,
the active regions move away from each other. Another important
physical process, flux injections, can be interpreted in terms of
the variation of the parameter, $\sigma$, in Equation (\ref{BCs}).
This kind of variations do not change the shape of $\Theta(\mu)$,
but change the magnitude of flux.
For instance, if an opposite polarity magnetic flux is injected
from below, due to the magnetic cancellation with the pre-existed
magnetic flux at the magnetar surface, the absolute value of the
parameter $|\sigma|$ might decrease. Some interesting consequences
from both kinds of alterations will be explored in this paper (see
Section 4).

\subsection{Inverse and Normal Configurations}
The solution to Equation (\ref{inhomoGS}) associated with the
boundary condition (\ref{BCs}) is of vital importance for our
further discussion. With some trivial boundary conditions, this
equation can be solved analytically by the Green-function method
(Lin et al. 1998). When more complex multipolar boundary
conditions, such as Equations (\ref{BCs}) and
(\ref{thetafunction}), are introduced, this GS equation can
generally be solved by the variable separation method. In this
paper, we develop a numerical method to solve this equation and
full solutions to this inhomogeneous GS equation are given in
details in Appendix B.

When solving the GS Equation (\ref{inhomoGS}) together with the
boundary condition Equations (\ref{BCs}) and
(\ref{thetafunction}), we find that there exist two kinds of
magnetic field configurations in the magnetosphere.
One is the the normal configuration, which means that the magnetic
field at the position ($r$, $\theta$) = ($h-r_0$, $\pi/2$) threads
across the flux rope in the same direction as the magnetar surface
magnetic field underneath at the equator ($r$, $\theta$) = ($r_s$,
$\pi/2$). The other is the inverse configuration, in the sense
that the magnetic field at the position ($r$, $\theta$) =
($h-r_0$, $\pi/2$) threads across the flux rope in the opposite
direction to the magnetar surface magnetic field underneath at the
equator ($r$, $\theta$) = ($r_s$, $\pi/2$). In our calculations,
the current $J$ is always kept positive\footnote{This is to avoid
the negative values of flux rope minor radius $r_0$, according to
Equation (\ref{internal}).}. So for our particular choice of
boundary conditions, if $\sigma$ is negative, we will get an
normal configuration, if $\sigma$ is positive, an inverse one is
obtained. Two schematic figures, both inverse and normal, are
shown in the left and right panel of Fig. 2, respectively.


\begin{figure}
\begin{center}$
\begin{array}{cc}
\includegraphics[width=3.2in]{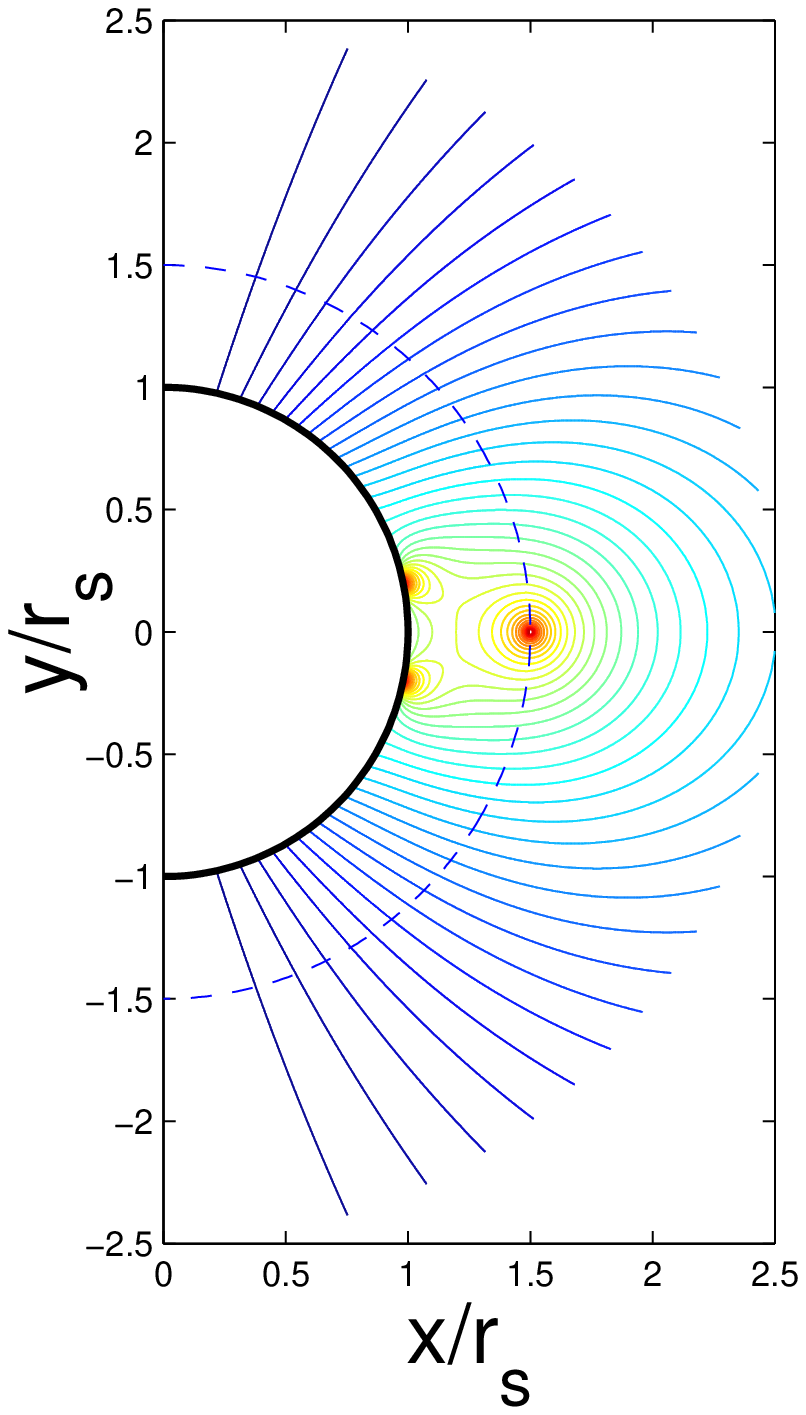} & \includegraphics[width=3.2in]{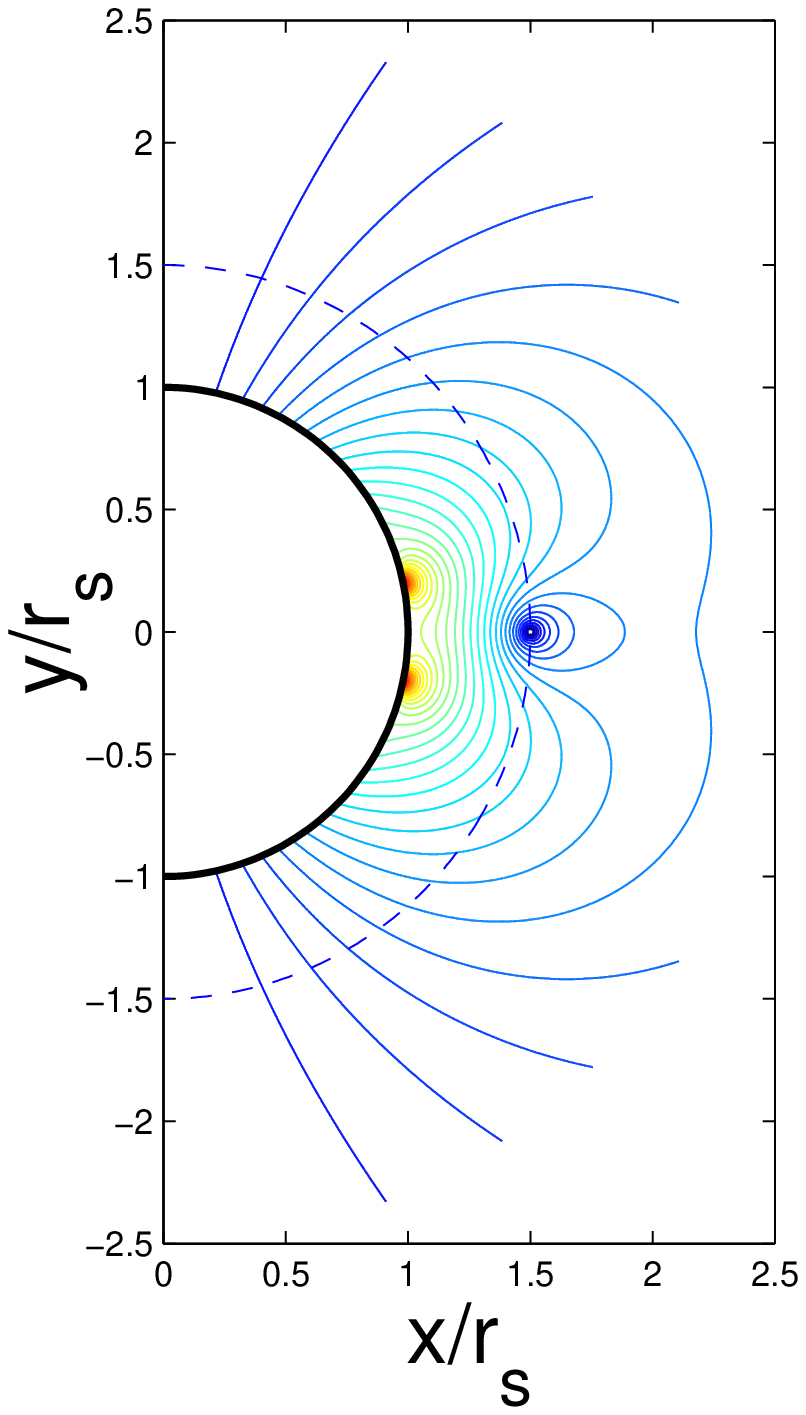} \\
\end{array}$
\end{center}
\caption{Magnetic field lines for both the inverse (left panel)
and normal (right panel) magnetic field configurations. Field
lines are obtained with solutions of the inhomogeneous GS equation
(see Appendix B). The thick black line denotes the magnetar
surface. The dashed line is a circle with a radius $r=h$, where
$h$ is the height of the flux rope. The flux rope lies at the
position ($r$, $\theta$) = ($h$, $\pi/2$).
At the magnetar surface $r=r_s$, two additional active regions
appear due to our choice of boundary conditions. }
\end{figure}


\section{Local-Internal and Global-External Equilibrium Constraints}
In what follows, we consider that the magnetar magnetosphere
evolves on a sufficiently long timescale so that we can treat the
magnetosphere as being essentially in a quasi-static equilibrium.
The condition for the flux rope equilibrium includes two parts:
the local-internal equilibrium and global-external equilibrium
(Forbes \& Isenberg 1991).
\subsection{Local-Internal Equilibrium Constraints}
For the local internal equilibrium, we assume that the force-free
condition, $\mathbf{J}\times \mathbf{B} = 0$, also applies within
the flux rope. We adopt the Lundquist (1950) force-free solution
to represent the distribution of current density and field inside
the flux rope\footnote{The Lundquist solution is obtained in
cylindrical coordinates (see Appendix A). Strictly speaking, the
Lundquist solution is only applicable for a straight cylindrical
twisted flux rope.}. For a circular toroidal flux rope in our
case, the Lundquist solution inside the flux rope is still valid
as long as the minor radius $r_0$ is much smaller than the major
radius, $h$ (also know as the flux rope height). In this case, a
simple relation between the minor radius of the flux rope $r_0$
and the current flowing in the flux rope, $I$, can be written as
\begin{equation}\label{internal}
r_0 =  \frac{r_{00} I_0}{I} = \frac{r_{00}}{J} \ ,
\end{equation}
where $J$ is the dimensionless current scaled by $I_0$ and
$r_{00}$ is the value for $r_0$ as $J=1$. We take $r_{00} = 0.01$
throughout this paper.
It should be noted that the internal equilibrium constraint,
Equation (\ref{internal}), is actually an alternative way
indicating the conservation of axial magnetic flux inside the flux
rope (see Appendix A).

\subsection{Global-external Equilibrium Constraints}
The global equilibrium is satisfied when the total force exerted
on the flux rope vanishes. The ring current inside the flux rope
provides an outward force.
Intuitively, the anti-parallel orientation of the current flowing
on the opposite sides of the ring produces a repulsive force
similar to the force between two parallel wires with anti-parallel
currents does.
The magnitude of this force is equal to the current, $I$, times
the magnetic field $B_{\mathrm{s}}$ (Shafranov 1966):
\begin{equation}
B_{\mathrm{s}} = \frac{I}{c h}\left(\ln\frac{8 h}{r_0} - 1\right)
\ ,
\end{equation}
where $r_0$ is the minor radius of the toroidal flux rope. The
additional terms in the bracket of the above equation appear due
to the curvature effects of the circular ring current
\footnote{For two straight wires, terms in the bracket disappear
and the induced magnetic field is strictly proportional to the
electric current and inversely proportional to the distance
between the wires.}. This ring current induced force must be
balanced by the external field $B_\mathrm{e}$.
The external magnetic field $B_\mathrm{e}$ at $r=h$ and $\theta =
\pi/2$ can be written explicitly as (The contribution from the
current inside the flux loop must be excluded for the external
magnetic field $B_{\mathrm{e}}$. For details, see Appendix B)
\begin{equation}\label{extB}
B_{\mathrm{e}} = \sum_{n \ \mathrm{odd}} n \Gamma_n d_n h^{-n-2} \ , %
\end{equation}
where the coefficients $\Gamma_n$'s and $d_n$'s are all given
explicitly in Appendix B.
The mechanical equilibrium condition, which requires matching the
external field $B_\mathrm{e}$ with $B_\mathrm{s}$, reads
\begin{equation}
f(\sigma, J, h) = 0 \ ,
\end{equation}
where
\begin{equation}\label{Bmatch}
f(\sigma, J, h) \equiv \left[ \sum_{n \ \mathrm{odd}}  n \Gamma_n
d_n h^{-n-2} \right]
 - \frac{J}{h}\left(\ln\frac{8Jh}{r_{00}} - 1 \right) \ .
\end{equation}
Note that the local-internal equilibrium constraint has also been
exploited in this equation. 
Note that the scaled current $J = I/I_0$ (not the current $I$)
appears in this equation. This is because all quantities, such as
$h$, $d_n$, and $I$, in this equation are measured in the
dimensional units mentioned above (see Section 2.2 and also
Appendix B).

The stream function $\Psi$ satisfies the ideal frozen-flux
condition, which provides a link between the electric current
flowing inside the flux rope and the boundary conditions at the
neutron star surface. Specifically, it requires that the stream
function on the edge of the flux rope remain constant as the
system evolves. At the equator $\theta = \pi/2$, the edge of the
flux rope is located at $r = h-r_0$ and the frozen-flux condition
can be written explicitly as
\begin{equation}
\Psi\left(h-r_0, \frac{\pi}{2}\right) = \mathrm{const} \ ,
\end{equation}
where $h$ and $r_0$ are the major and minor radius of the flux
rope, respectively. Substituting $r= h-r_0$ and $\theta = \pi/2$
into the stream function, i.e., Equation (B1) in Appendix B, we
arrive at another constraint,
\begin{equation}
g(\sigma , J, h) = \mathrm{const} \ ,%
\end{equation}
where
\begin{equation}\label{frozen}
g(\sigma , J, h) \equiv \sum_{n \ \mathrm{odd}} \left[ \Gamma_n  c_{n} \left(1-\frac{r_{00}}{Jh}\right)^{n+1} %
+ \Gamma_n d_{n} \left(h-\frac{r_{00}}{J}\right)^{-n} \right]  \ . %
\end{equation}
where $d_n$'s and $\Gamma_n$'s have the same meaning as those in
Equation (\ref{Bmatch}), $c_n$'s are explicitly given in Appendix
A.
Note again that all quantities in this equation are measured in
the dimensional units defined above. The quantity $r_{00}$ appears
because of the internal equilibrium constraint.

In summary the equilibrium constraints, including the force
balance and the frozen-flux condition, can be written as the
following form
\begin{equation}\label{eqm}
 \left\{ \begin{array}{lll}
f(\sigma, J, h) = 0 \   \\
g(\sigma, J, h) = \mathrm{const} \   \\
\end{array} \right.  \ ,
\end{equation}
where functions $f$ and $g$ are defined in Equations
(\ref{Bmatch}) and (\ref{frozen}). Numerical values of $f$ and $g$
are calculated following the procedures presented in Appendix B.
For a given value of $\sigma$, the above two equations could be
treated as a nonlinear set of equations for $J$ and $h$, which can
be solved numerically by the Newton-Raphson method (Press et al.
1992).
Note that the Jacobi matrix necessary for the Newton-Raphson
method is hard to obtain analytically and we calculate the Jacobi
matrix numerically instead.

\section{Loss of Equilibrium in Response to Variations at Magnetar Surface}
We consider the possibility that the primary mechanism for driving
a magnetar giant flare is a catastrophic loss of equilibrium. The
loss of equilibrium is initiated by
slow changes at the magnetar surface. Physically, there are
generally two possible processes that could occur at the magnetar
surface.
One is that new magnetic fluxes, driven by the plastic deformation
of neutron star crust, may be injected continuously into the
magnetosphere (Klu\'{z}niak \& Ruderman 1998; Thompson et al.
2002; Lyutikov 2006; G\"{o}tz et al. 2007). Another interesting
possibility is brought about by the crust horizontal movement
(Ruderman 1991; Jones 2003). It is very difficult to compress
magnetar crust material very much, or to move elements of crust up
or down. It is, however, much easier to move parts of the crust
horizontally, in ways which apply only shear strains to it
(Thompson \& Duncan 2001; Jones 2003). It is possible that, when
the magnetic field is strong enough, the interior magnetic stress
may cause the active regions of the crust to move horizontally
(Ruderman 1991).

For the first possibility, as the new current-carrying magnetic
fluxes are injected, a direct consequence is that the background
magnetic field would vary gradually because of the active flux
injections prior to large outbursts. The background magnetic field
would increase (decrease) if the same (opposite) polarity flux is
injected.
Variations of the equilibrium height of flux rope with alterations
in the background magnetic field are carefully examined for both
the inverse and normal magnetic configurations. In this case, we
fix the value of $\mu_0 = 0.1$ and investigate the effects of
variations of $\sigma$ on the flux rope equilibrium height.
Numerical results of Equation (\ref{eqm}) are shown in Fig 3. and
Fig 4. These two figures show the results for the normal and
inverse magnetic configurations, respectively. The curves in Fig.
3 and Fig. 4 consist of two branches, which comes from the fact
that there exist two roots of $h$ for each particular value of
$\sigma$ and the two roots lie on separate branches. The upper
branch denotes an unstable equilibrium state
because when the equilibrium is on the upper branch, an slight
upward vertical displacement will generate an outward driving
force.
The lower branch, however, stands for a stable branch, in the
sense that an slight upward displacement would create an inward
restoring force, just like an harmonic oscillator. The stability
of the flux rope could be understood in terms of a spring model.
The spring coefficient of Hooke's law determines the stability of
the spring. It would be instructive to treat the total force, $T$,
as a function of flux rope height, $h$, while keeping the flux
frozen condition satisfied. Detailed analysis shows that
derivative $dT/dh$ (which is equivalent to the spring coefficient
of Hooke's law) is negative if the flux rope lies on the lower
branch and positive on the upper branch (see Fig. 6.18 in Forbes
2010). Negative $dT/dh$ corresponds to a normal Hooke spring
coefficient and a stable equilibrium, while positive $dT/dh$
corresponds to an anomalous Hooke spring coefficient and an
unstable equilibrium. The two branches are connected by a critical
point (nose point). The instability threshold lies at the nose
point. The nose point can also be understood as the critical
loss-of-equilibrium point (red point in Fig. 3 and Fig. 4).
Once the equilibrium reaches the loss-of-equilibrium critical
height, the system would no longer stay in a stable equilibrium
state. The flux rope will lose equilibrium and lead to an
eruption.
In the case of normal configuration, Fig. 3 shows that the
increase of the parameter $|\sigma|$ (flux injection of the same
polarity) would bring the system to the critical point, which
means that the enhancement of the background flux would trigger
the catastrophic behavior for a normal configuration. While for
the case of inverse configuration, Fig. 4 shows the decrease of
the parameter $|\sigma|$ (flux injection of the opposite polarity)
would lead to loss of equilibrium, indicating the decay of the
background flux works for an inverse configuration. Our
calculations show that the critical height for the two kind of
configurations differs much. The normal configuration shows a
rather low critical height, roughly 3\% percent above the magnetar
surface, which is, for a typical neutron with radius $10^6$cm,
roughly $3\times10^4$cm. While for the inverse configuration, the
critical height is about 20\% above the magnetar surface, which is
about $2\times10^5$cm.
Given the regular arrangements that occur at the magnetar surface,
the small critical height of the normal configuration would
indicate that it may not survive those arrangement at the magnetar
surface and the inverse configuration, whose critical height is
larger, is preferred in real circumstances.

\begin{figure}
\begin{center}
\epsfig{file=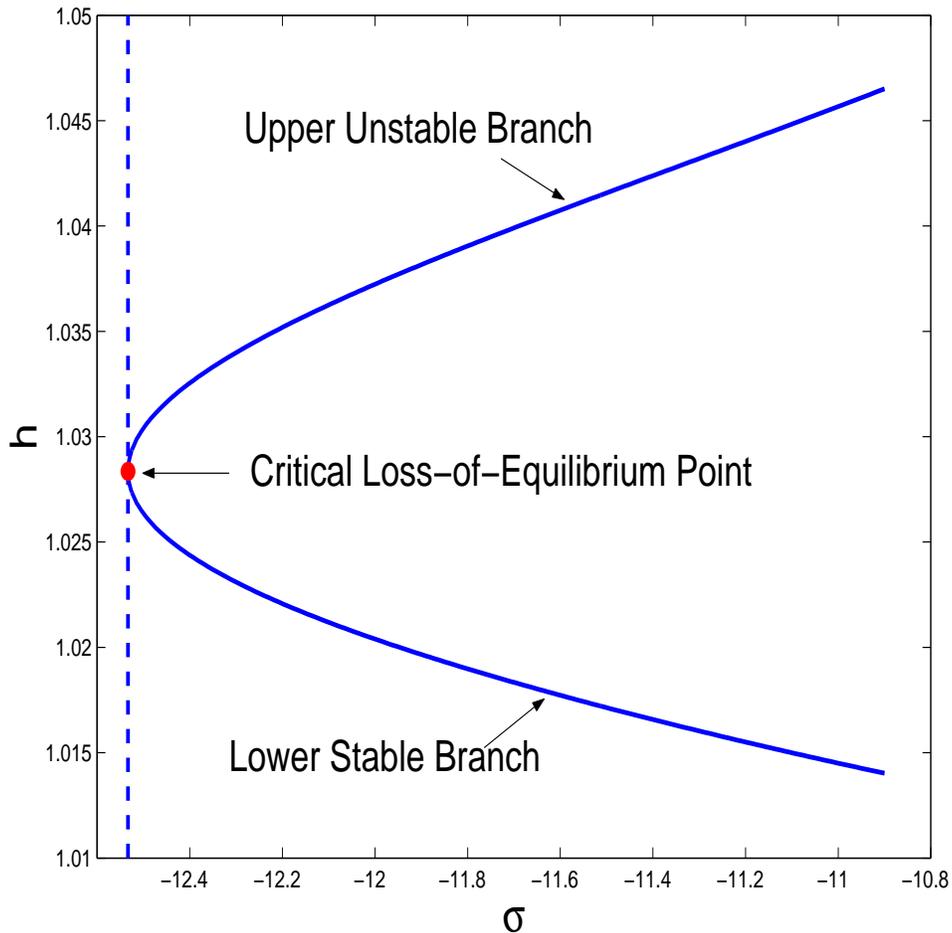,height=5in,width=5in,angle=0}
\end{center}
\caption{
         Equilibrium height $h$ as a function of $\sigma$.
         This curve is numerically obtained as solutions to
         Equation (\ref{eqm}).
         We always choose the current to be positive.
         The negative value of $\sigma$ denotes a normal
         magnetic configuration. In this case, with the
         increase of of $|\sigma|$, the equilibrium height
         gradually increases and reaches the critical loss-of-equilibrium
         point, beyond which the flux rope could not maintain
         the stable equilibrium. The critical height for the
         normal magnetic configuration is approximately
         $h_c = 1.028$. All lengths are scaled by $r_s$.
         }
\label{sigmahnormal}
\end{figure}

\begin{figure}
\begin{center}
\epsfig{file=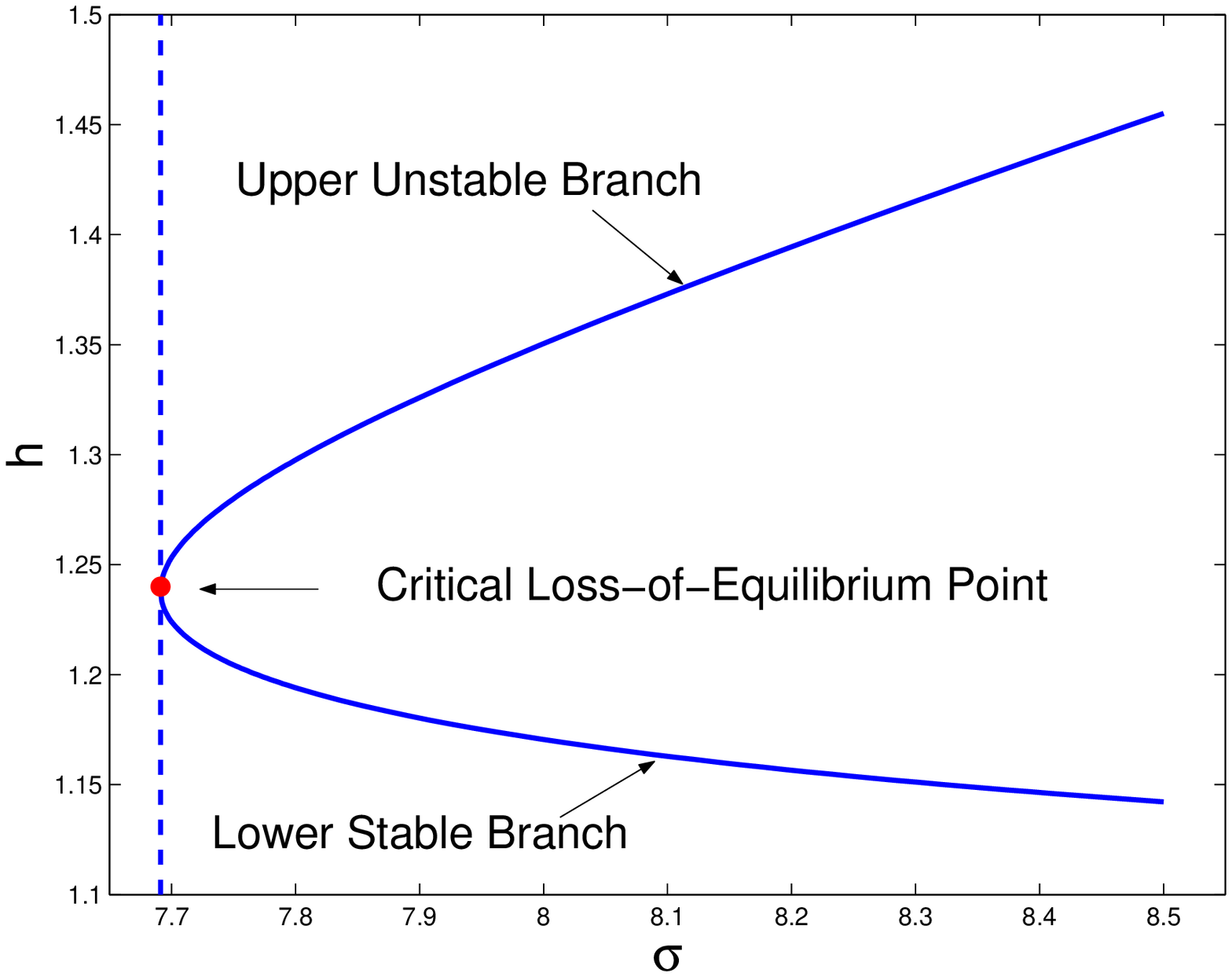,height=5in,width=5in,angle=0}
\end{center}
\caption{
         The same as Fig. 3 but for an inverse magnetic field
         configuration. In this case, with the
         decrease of of $|\sigma|$, the equilibrium height
         gradually increases and reaches the critical loss-of-equilibrium
         point, beyond which the flux rope could not maintain
         the stable equilibrium. The critical height for
         normal magnetic configuration is approximately
         $h_c = 1.24$. All lengths are scaled by $r_s$.
         }
\label{sigmahinverse}
\end{figure}




As the interior magnetic stress by the ultra-strong magnetic field
in magnetar may cause active regions of the crust to move
horizontally (Ruderman 1991; Jones 2003; Lyutikov 2006),
the relative positions between multipolar active regions may vary,
moving apart or approaching each other, which may also have
significant implications for the catastrophic behavior of
magnetospheres. Consequently, we further investigate the response
of the flux rope to horizontal motions of active regions at the
magnetar surface.
In our simplified model, the distance between two active regions
is determined by a single parameter $\mu_0$ in Equation
(\ref{thetafunction}).
The increase of $\mu_0$ may indicate that active regions move
apart (see Fig.1). To investigate the effects of horizontal
motions, we fix the value of $\sigma$ and vary the parameter
$\mu_0$. Again, we numerically solve Equation (\ref{eqm}) and get
two roots of $h$ for each particular value of $\mu_0$. We show in
Fig. 5, taking the inverse configuration as an example, the
variation of the equilibrium height with the distance between two
active regions. Similar to Fig. 3 and Fig. 4, the upper branch and
the lower branch denotes unstable and stable equilibrium,
respectively. As the two active regions move apart, the
equilibrium height gradually increases and reaches the critical
height $h_c=1.32$ when $\mu_0 = 0.14$. After reaching the critical
point, the system would no longer maintain a stable equilibrium
state. This means that, in addition to flux injections, the
horizontal motions of the active regions could give rise to the
loss of equilibrium and dynamical eruptions as well.


\begin{figure}
\begin{center}
\epsfig{file=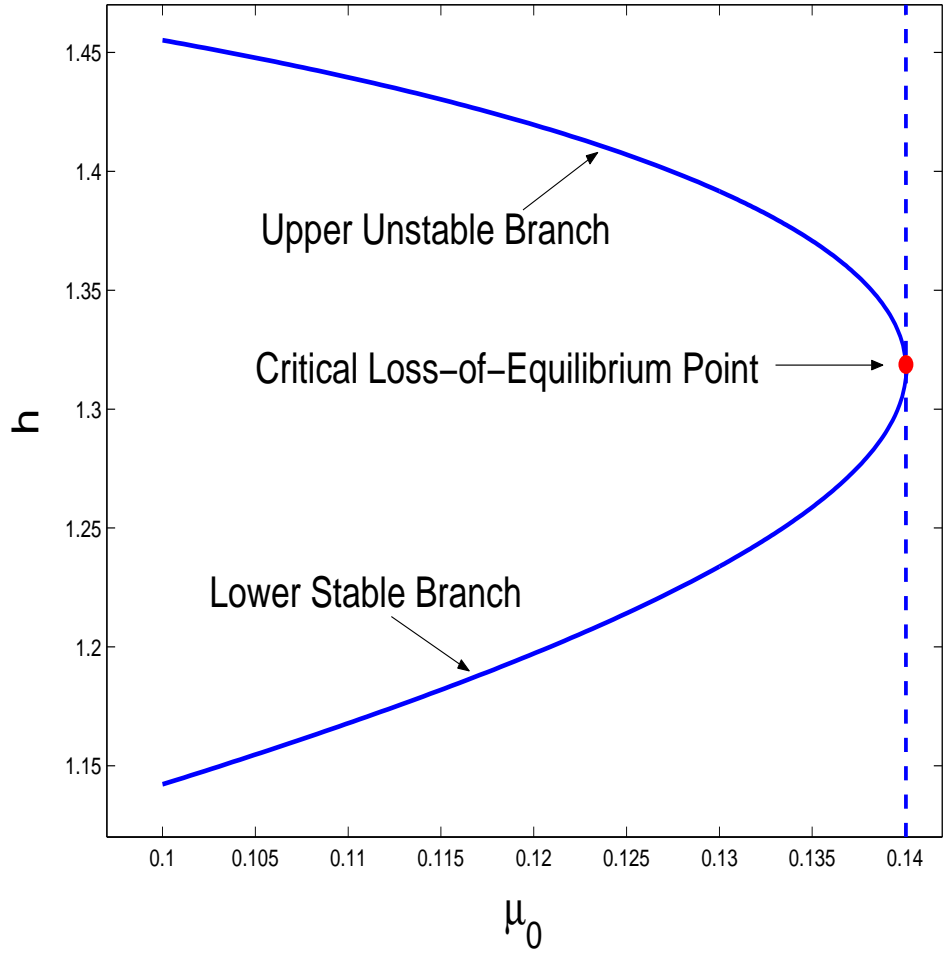,height=5in,width=5in,angle=0}
\end{center}
\caption{
         Equilibrium height $h$ as a function of $\mu_0$.
         We fix $\sigma = 8.5$ in this figure.
         With the increase of of $\mu_0$,
         the two active regions move apart. The equilibrium height
         gradually increases and reaches the critical point, beyond
         which no stable equilibrium state exists. The critical height for
         normal magnetic configuration is approximately
         $h_c = 1.32$. All lengths are scaled by $r_s$.
         }
\label{mufig}
\end{figure}


\section{Conclusions and Discussions}\label{sec:diss}
In this work, we consider the possibility that the primary
mechanism for driving an eruption in magnetar giant flare is a
catastrophic loss of equilibrium of a helically twisted flux rope
in the magnetar magnetosphere. The loss of equilibrium behavior of
a flux rope is investigated in a multi-polar magnetic field
configuration, taking into account possible effects of flux
injections and crust horizontal motions. The loss of equilibrium
model describes a quasi-static equilibrium that varies in response
to slow changes at the magnetar surface. Beyond a critical point,
the stable equilibrium can not be maintained and the transition to
a dynamical evolution naturally occurs.

Equilibrium states of a stationary, axisymmetric magnetic field in
the non-rotating magnetosphere containing a flux rope are obtained
as solutions of the inhomogeneous GS equation in a spherical polar
coordinate. In view of the complex multipolar boundary conditions
at the magnetar surface, we develop a numerical method to solve
the GS equation. Two kinds of magnetic field configurations,
inverse and normal, are carefully examined in this work. Both of
them present the loss of equilibrium behavior. We carefully
examined the critical height of the flux rope beyond which a
stable equilibrium could not be maintained and a sudden release of
magnetosphere will be triggered. We find that the critical flux
rope height is different for the two types of configurations. We
also investigate effects of another form of boundary changes,
crust horizontal motions, on the loss of equilibrium behavior. Our
results show that both the flux injection and crust motions could
trigger the catastrophic behavior in the magnetosphere.





In our simplified model, the flux rope is assumed to be a closed
current ring encircling the magnetar. It is suspended in the
magnetar magnetosphere and two ends of the flux rope are not
anchored to the magnetar surfaces.
We expect that the overall catastrophic behavior of our model
should remain the same even the anchoring effects of flux rope is
taken into account. In order to further understand the anchoring
effects on the catastrophic behavior, more realistic three
dimensional model that includes a flux rope with two ends anchored
to the magnetar surface is worth further investigation.

The magnetic energy that could be released in our model is an
interesting issue that is worthwhile to be explored further.
For a spherical coordinate in our paper, the ground energy
reference state, based on which the fraction of released magnetic
energy can be calculated, is the fully open Aly-Sturrock field
(Aly 1984, 1991; Sturrock 1991; Yu 2011b). Since the boundary
conditions in our paper is more complex (not dipolar or
quadrupolar boundary conditions), the construction of the
Aly-Sturrock field is technically nontrivial. As a result, the
fraction of the energy that could be released in our model
involves a very careful calculation of the Aly-Sturrock field.
Intuitively, according to prior studies, which have shown that
magnetic configurations with more complex boundary conditions
would be able to release more energies (Forbes \& Isenberg 1991;
Isenberg et al. 1993), as well as the complex boundary conditions
adopted in our model, it is expected that our model would release
enough energy for a magnetar giant flare\footnote{Typically 1\% of
the magnetic energy release could already account for a giant
flare. For a simple dipolar boundary condition, about 1\% of the
magnetic energy can be released (Forbes \& Isenberg 1991).
However, about 5\% of the magnetic energy can be released for a
quadrupolar boundary condition (Isenberg et al. 1993).}. Full
details of the energetics of our model would be discussed
elsewhere.

It is possible that the current sheet forms after the system loses
equilibrium (Forbes \& Isenberg 1991; Forbes \& Priest 1995). With
the formation of current sheet, the tearing instability would
develop inside the current sheet (Komissarov et al. 2007) and the
subsequent magnetic reconnection would further accelerate the flux
rope (Priest \& Forbes 2000).
Magnetic field configurations with the current sheet in a
spherical polar coordinate is a long-standing unresolved problem.
We note that numerical method developed in this work can be
further extended to allow the presence of current sheets (Yu
2011b). Further discussions about the current sheets formation and
their effects on the catastrophic behavior would be reported in a
separate paper.

It would be interesting to to examine the spectral properties of
the model presented in this paper (Thompson et al 2002),
in which a locally twisted flux rope is self-consistently
incorporated into the magnetar magnetosphere. By fitting the
spectral features with observations (Pavan et al. 2009), certain
parameters in this model, e.g., flux rope height, electric current
and magnetic field, may be better constrained. In parallel, recent
Fermi observation of Crab nebulae gamma-ray flare could possibly
be explained by the magnetic reconnection models (Abdo et al.
2011), in which the loss of equilibrium may be the trigger for the
formation of current sheet and subsequent magnetic reconnection
processes. The high energy flare emission from Crab Nebulae is
thought to be synchrotron radiation by relativistic
electron-positron pairs accelerated in this current sheet
(Uzdensky et al. 2011). It would be instructive to calculate,
based on the model shown in this paper, the emission spectra and
compare them with recent Crab Nebula observations.


The models constructed in this work are likely to be useful as
initial states in high resolution force-free electrodynamic
numerical simulations to explore the dynamics of magnetic
eruptions (Yu 2011a). Our current model can not address the
dissipation processes that occur during giant flares. This is left
for a future work to directly simulate the behavior of loss of
equilibrium and relevant dissipation processes using a newly
developed resistive force-free electrodynamic code (Yu 2011a).



\acknowledgments

Discussions with T. Forbes \& J. Lin are highly appreciated. We
are grateful to the anonymous referee's insightful comments that
improve this paper. The research is supported by the Natural
Science Foundation of China (grants 10703012, 11173057), the
Western Light Young Scholar Program. The computation in this work
is performed at HPC Center, Kunming Institute of Botany, CAS,
China.

\clearpage


\appendix
\section{Lunquist's solution Interior to Flux Ropes}
When the flux rope minor radius is small compared to the major
radius of the flux rope, the flux rope interior magnetic field
could be approximated as a straight force-free magnetic cylinder.
We use Lundquist's solution to represent the flux rope interior
magnetic field. The solution satisfies the force-free condition
$\nabla\times\mathbf{B}$ = $\lambda \ \mathbf{B}$, where $\lambda$
is a constant. The solution can be written in a cylindrical
coordinate $(r, \theta ,z)$. Note that the $z$ direction in this
Appendix is actually the $\phi$ direction in the main text.
Explicitly, Lundquist's solution is
\begin{equation}
B_r = 0 \ ,
\end{equation}
\begin{equation}
B_{\theta} = B_0 J_1(\lambda r) \ ,
\end{equation}
\begin{equation}
B_{z} = B_0 J_{0}(\lambda r) \ ,
\end{equation}
where $B_z$ is along the central axis, $B_{\theta}$ is the
azimuthal component, and $B_r$ is the radial component, $J_0$ and
$J_1$ are the zeroth- and first-order Bessel functions. The
conservation of the toroidal flux simply gives
\begin{equation}
\int^{2\pi}_0 \int^{r_0}_0 B_z r drd\theta = \mathrm{const} \ .
\end{equation}
Substituting the force-free condition $\lambda B_z = 4\pi j_z/c$,
we have that
\begin{equation}
\frac{4\pi}{\lambda c}\int^{2\pi}_0 \int^{r_0}_0 j_z r drd\theta =
\frac{4\pi I}{\lambda c} = \mathrm{const} \ ,
\end{equation}
which can be written equivalently as
\begin{equation}
\frac{I}{\lambda} = \mathrm{const} \ .
\end{equation}
At the surface of the flux rope, $B_z$ is zero so $J_0({\lambda
r_0}) = 0$. Therefore $\lambda r_0$ is the first zero of $J_0$ and
$\lambda r_0 = 2.405$. We can finally arrive at
\begin{equation}
r_0 I = \mathrm{const} \ ,
\end{equation}
which is Equation (\ref{internal}) in the main text.

\section{General Solution of the inhomogeneous
Grad-Shafranov Equation with Multipolar Boundary Conditions}

According to the variable separation method, the general solution
to the Grad-Shafranov equation can be written as (see Yu 2011b)
\begin{equation}
\Psi = \sum_{n \ \mathrm{odd}} %
\left( c_{n} R_{n}(r) + d_{n} r^{-n} \right) %
\left[ \frac{P_{n-1}(\mu) - P_{n+1}(\mu)}{2n+1} \right] \ , \ %
\mathrm{n = odd} \ ,
\end{equation}
where $c_n$ and $d_n$ are constant coefficients to be specified,
$P_{n-1}(\mu)$ and $P_{n+1}(\mu)$ are Legendre polynomials, $\mu =
\cos\theta$. Here the piece-wise continuous function $R_n(r)$ in
the above equation is defined as
\begin{equation}
R_{n}(r) = \left\{ \begin{array}{ll}
\left( r/h \right)^{n+1} & r \leq h \\
\left( h/r \right)^{n}   & r \geq h \\
\end{array} \right.  \ .
\end{equation}
Note that the derivative of the function $R_n(r)$ is
discontinuous. This feature is exploited in the following to
handle the inhomogeneous source terms associated with the
Dirac-$\delta$-type current density. Also note an identity for the
Legendre polynomials
\begin{equation}
\frac{P_{n-1}(\mu) - P_{n+1}(\mu)}{2n+1} = \frac{(1 -
\mu^2)}{n(n+1)}\frac{d P_n}{d \mu} \ .
\end{equation}
The inhomogeneous Grad-Shafranov (GS) equation reads
\begin{equation}
\frac{\partial^2 \Psi}{\partial r^2} %
+ \frac{\sin\theta}{r^2}\frac{\partial}{\partial \theta}%
\left( \frac{1}{\sin\theta}\frac{\partial \Psi}{\partial\theta} \right) %
 = -r\sin\theta \frac{4\pi}{c} J_{\phi} \ .
\end{equation}
Substituting Equation (B1) and the explicit expression of
$J_{\phi}$ (Equation (\ref{currentdensity})) into the
inhomogeneous GS equation, we arrive at
\begin{equation}
\sum_{n} c_n \frac{(1-\mu^2)}{n(n+1)} \frac{d P_n}{d \mu} \left[
\frac{d^2 R_n}{d r^2} - n(n+1)\frac{R_k}{r^2} \right] = - r
\sin\theta \ \frac{4\pi I}{h c}\ \delta(\cos\theta) \delta(r-h) \
.
\end{equation}
Integrating $r$ over an infinitesimally thin shell around $r=h$,
we can rewrite the above equation as \footnote{Note the fact that
the first order derivative of $R_n(r)$ is discontinuous, and
$\frac{d R_n}{dr}\big|_{r=h^+} - \frac{d R_n}{dr}\big|_{r=h^-} =
-(2n+1)\frac{1}{h}$.}
\begin{equation}
\sum_{n} c_n \frac{(1-\mu^2)}{n(n+1)} \frac{d P_n}{d\mu} \left[
-(2n+1)\frac{1}{h} \right] = - \frac{4\pi I}{c} \sin\theta
\delta(\cos\theta) \ .
\end{equation}
Multiplying $\sin\theta \frac{dP_n}{d\mu}$ on both sides of the
above equation and integrating $\theta$ over $[0, \pi]$ , we have
that
\begin{equation}
c_n = %
\left[ \frac{(-1)^{\frac{n-1}{2}} n!}{2^{n}\left(\frac{n-1}{2}\right)!  %
\left(\frac{n-1}{2}\right)!} \right] \frac{4 \pi I h}{c} \ . %
\end{equation}
It is more convenient to calculate $c_n$ numerically by the
following recursive relation,
\begin{equation}
c_1 = 0.5 \ \left( \frac{4 \pi I h }{c} \right) \ , \ c_{n+2} = -
\left( \frac{n+2}{n+1} \right) \ c_n \ .
\end{equation}
Note that terms in the stream function involving $c_n$'s are
induced by the current inside the flux rope 
and the external magnetic field in Equation (\ref{extB}) only
involves terms related to $d_n$'s in Equation (B1).
According to Equation (\ref{brbt}), the external magnetic field
$B_{\mathrm{e}}$ at $r = h$ and $\theta = \pi/2$, can be written
explicitly as
\begin{equation}
B_\mathrm{e} = \sum_{n \ \mathrm{odd}} n \Gamma_n d_n h^{-n-2} \ , %
\end{equation}
where
\begin{equation}
\Gamma_n = \left[ \frac{P_{n-1}(\mu = 0) - P_{n+1}(\mu = 0)}{2n+1}
\right] \ .
\end{equation}
The coefficient $\Gamma_n$ can be readily calculated as
\begin{equation}
\Gamma_{1} = 0.5 \ , \ \Gamma_{n+2} = - \left( \frac{n}{n+3}
\right) \Gamma_{n} \ .
\end{equation}
To specify the coefficients of $d_n$'s, we have to take into
account the boundary conditions at the magnetar surface. We
require that the stream function $\Psi$ be equal to $\Psi_{s}$,
i.e., the boundary conditions (\ref{BCs}) at the neutron star
surface $r=r_s$.
To achieve this, we expand $\Psi_{s}(\mu)$ as the following
\[
 \Psi_s(\mu) = \sum_{n} a_n \left[ \frac{P_{n-1}(\mu) - P_{n+1}(\mu)}{2n+1}
 \right] \ .
\]
The coefficients $a_n$'s can be expressed as
\[
a_n = \frac{2n+1}{2} \Psi_0 \sigma \int^{1}_{-1} \Theta(\mu) \
\frac{d P_n(\mu)}{d \mu} d\mu \ .
\]
Finally the coefficients $d_n$'s can be obtained as follows,
\begin{equation}
d_n = r_s^n \left[ a_n - c_n \left(\frac{r_s}{h}\right)^{n+1}
\right] \ . \
\end{equation}
For numerical conveniences,  we scale all lengths by $r_s$, the
magnetic flux by $\Psi_0$ and the current by $I_0 = \Psi_0 c /r_s$
throughout this paper. Then the equation $B_{\mathrm{e}} =
B_{\mathrm{s}}$ becomes
\begin{equation}
f(\sigma, J, h) = 0 \ ,
\end{equation}
where
\begin{equation}
f(\sigma, J, h) \equiv \sum_{n \ \mathrm{odd}} n \Gamma_n d_n
h^{-n-2}
 - \frac{J}{h}\left(\ln\frac{8Jh}{r_{00}} - 1 \right) \ ,
\end{equation}
and $ J = I / I_0 \ $. Similarly the frozen-flux condition can be
written as
\begin{equation}
g(\sigma , J, h) \equiv \sum_{n \ \mathrm{odd}} %
\Gamma_n  c_{n} \left(1-\frac{r_{00}}{Jh}\right)^{n+1} %
+ \Gamma_n d_{n} \left(h-\frac{r_{00}}{J}\right)^{-n} = \mathrm{const} \ . %
\end{equation}
Here we can see that both $c_n$'s and $d_n$'s are explicitly
specified and the solution to the inhomogeneous GS equation
associated with the multipolar boundary conditions is uniquely
determined. The results in this Appendix establish the basis for
further investigations of the evolution of the whole system in the
main text.





\begin{thebibliography}{99}

\bibitem[Abdo 2011]{Abodo2011}
Abdo, A. A., et al. 2011, Science, 331, 739

\bibitem[Aly (1984)]{a84}
Aly, J. J., 1984, ApJ, 283, 349

\bibitem[Aly (1991)]{a91}
Aly, J. J., 1991, ApJL, 375, 61





\bibitem[Beloborodov{2009}]{Bdov09}
Beloborodov, A. M. 2009, ApJ, 703, 1044





\bibitem[Chen et al. 1997]{chen1997}
Chen, J., et al. 1997, ApJL, 490, 191


\bibitem[Dere et al. 1999]{Dere1999}
Dere, K. P., et al. 1999, ApJ, 516, 415

\bibitem[Duncan \& Thompson (1992)]{dt92}
Duncan, R. C., \& Thompson, C., 1992, ApJL, 392, 9


\bibitem[Feroci et al. 2001]{feroci01}
Feroci, M., Hurley, K., Duncan, R. C., \& Thompson, C. 2001, ApJ,
549, 1021

\bibitem[Forbes 2010]{Forbes2010}
Forbes, T. G. Heliophysics: Space Storms and Radiation: Causes and
Effects, 2010, Edited by Carolus, J. S. \& George, L. S.
(Cambridge Univ. Press), 159

\bibitem[Forbes \& Isenberg 1991]{Forbes91}
Forbes, T. G., \& Isenberg, P. A. 1991, ApJ, 373, 294

\bibitem[Forbes \& Priest 1995]{Forbes95}
Forbes, T. G., \& Priest, E. R. 1995, ApJ, 446, 377



\bibitem[Gaensler et al. 2005]{Gaensler05}
Gaensler, B. M., et al. 2005, Nature, 434,104

\bibitem[Gavriil et al. (2002)]{gav2002}
Gavriil, F. P., Kaspi, V. M., Woods, P. M. 2002, Nature, 419, 142


\bibitem[Gill \& Heyl 2010]{Gill2010}
Gill, R., \& Heyl, J. S. 2010, MNRAS, 407, 1926


\bibitem[G\"{o}tz et al. 2007]{Gotz07}
G\"{o}tz, D.,  Mereghetti, S., \& Hurley, K. 2007, Ap\&SS, 308, 51



%

\bibitem[Isenberg et al. 1993]{Isenberg93}
Isenberg, P. A., Forbes, T. G., \& D\'{e}moulin, P. 1993, ApJ,
417, 368

\bibitem[Jones 2003]{Jone2003}
Jones, P. B. 2003, ApJ, 595, 342


\bibitem[Kluzniak (1998)]{kluz98}
Klu\'{z}niak, W. \& Ruderman, M., 1998, ApJL, 505, 113-117


\bibitem[Komissarov et al. (2007)]{komi07}
Komissarov, S. S., Barkov, M. \& Lyutikov M., 2007, MNRAS, 374,
415


\bibitem[Kouveliotou(1998)]{kou98}
Kouveliotou, C., et al. 1998, Nature, 393, 235

\bibitem[Levin \& Lyutikov (2012)]{levin12}
Levin \& Lyutikov, arXiv:1204.2605

\bibitem[Lin et al. 1998]{lin98}
Lin, J., Forbes, T. G., Isenberg, P. A., \& D\'{e}moulin, P. 1998,
ApJ, 504, 1006




\bibitem[Lundquist 1950]{Lundquist1950}
Lundquist, S. 1950, Ark. Fys., 2, 361


\bibitem[Lyutikov (2006)]{lyutikov06}
Lyutikov, M. 2006, MNRAS, 367, 1602

\bibitem[Mazets et al. (1979)]{mazets79}
Mazets, E. P., et al. 1979, Nature, 282, 587


\bibitem[Mereghetti (2008)]{m08}
Mereghetti, S. 2008, A\&AR, 15, 225

\bibitem[Mereghette \& Stella (1995)]{ms95}
Mereghetti, S. \&  Stella, L. 1995, ApJL, 442, 17



\bibitem[Palmer et al. (2005)]{palmer05}
Palmer, D. M., et al. 2005, Nature, 434, 1107

\bibitem[Pavan et al. 2009]{pavan09}
Pavan, L., Turolla, R., Zane, S., \& Nobili, L. 2009, MNRAS 395,
753

\bibitem[Perna \& Gotthelf (2008)]{perna08}
Perna, R., \& Gotthelf, E. V. 2008, ApJ, 681, 522

\bibitem[Priest \& Forbes 2000]{priest2000}
Priest E., \& Forbes T. 2000, Magnetic Reconnection. MHD Theory
and Applications. Cambridge Univ. Press, Cambridge

\bibitem[Press et al. 1992]{press92}
Press, W. H., Teukolsky, S. A., Vetterling, W. T., \& Flannery, B.
P. 1992, Numerical Recipes in FORTRAN 2nd Ed., Cambrdige Univ.
Press, Cambridge


\bibitem[Ruderman 1991]{ruderman91}
Ruderman, M. 1991, ApJ, 366, 261

\bibitem[Shafranov (1966)]{s66}
Shafranov, V. D. 1966, Rev. Plasma Phys., 2, 103

\bibitem[Sturrock (1991)]{s91}
Sturrock, P. A. 1991, ApJ, 380, 655


\bibitem[Thompson \& Duncan (1996)]{td96}
Thompson, C., \& Duncan, R. C. 1996, ApJ, 473, 322

\bibitem[Thompson \& Duncan (2001)]{td01}
Thompson, C., \& Duncan, R. C. 2001, ApJ, 561, 980

\bibitem[Thompson et al. (2002)]{thompson2002}
Thompson, C., Lyutikov, M., \& Kulkarni, S. R. 2002, ApJ, 574, 332



\bibitem[Uzdensky (2011)]{uz2011}
Uzdensky, D. A., Cerutti, B. \& Begelman, M. C. 2011, ApJL, 737,
40

\bibitem[van Ballegooijen 1989]{van89}
van Ballegooijen, A. A., \& Martens, P. C. H. 1989, ApJ, 343, 971


%



\bibitem[Woods (2001)]{woods01}
Woods, P. M., et al., 2001, ApJ, 552, 748

\bibitem[Woods \& Thompson (2006)]{wt06}
Woods, P. M., \& Thompson, C., 2006, in Compact Stellar X-Ray
Sources, ed. W. H. G. Lewin \& van der Klis (Cambridge Univ.
Press), 547

\bibitem[Woods et al. (2007)]{woods07}
Woods, P. M., et al., 2007, ApJ, 654, 470

\bibitem[Yu (2011)]{yu2011a}
Yu, C., 2011a, MNRAS, 411, 2461

\bibitem[Yu (2011)]{yu2011b}
Yu, C., 2011b, ApJ, 738, 75


%

\end{thebibliography}

\end{document}